\documentclass[final,1p,times]{elsarticle} 

\usepackage{graphicx}
\usepackage{amssymb} 
\usepackage{amsthm} 
\usepackage{lineno}

\journal{Nuclear Physics A} 

\begin{document}

\begin{frontmatter} 

\title{The RHIC Beam Energy Scan Program: Results from the PHENIX Experiment}

\author{Jeffery T. Mitchell (for the PHENIX\fnref{col1} Collaboration)}
\fntext[col1] {A list of members of the PHENIX Collaboration and acknowledgements can be found at the end of this issue.}
\address{Brookhaven National Laboratory, Building 510C, P.O. Box 5000, Upton, NY 11973-5000 USA}

\begin{abstract} 
The PHENIX Experiment at RHIC has conducted a beam energy scan at several collision energies in order to search for signatures of the QCD critical point and the onset of deconfinement. PHENIX has conducted measurements of transverse energy production, muliplicity fluctuations, and the skewness and kurtosis of net charge distributions. The data analyzed to date show no significant indications of the presence of the critical point.
\end{abstract} 

\end{frontmatter} 


\section{Introduction}

Recent lattice QCD calculations predict that there is a first order phase transition from hadronic matter to a Quark-Gluon Plasma that ends in a critical point, with a continuous phase transition on the other side of the critical point. The Relativistic Heavy Ion Collider (RHIC) has conducted a program to probe different regions of the QCD phase diagram in the vicinity of the possible critical point with a beam energy scan. During 2010 and 2011, RHIC provided Au+Au collisions to PHENIX at $\sqrt{s_{NN}} = $ 200 GeV, 62.4 GeV, 39 GeV, 27 GeV, 19.6 GeV, and 7.7 GeV.  Analysis of the data concentrates on two strategies: looking for signs of the onset of deconfinement by comparing to results at the top RHIC energy, and searching for direct signatures of a critical point. Results from PHENIX covering transverse energy production, multipicity fluctuations, and the skewness and kurtosis of net charge distributions will be discussed in this article.

\section{Transverse Energy Production}

\begin{figure}[htbp]
\begin{center}
\includegraphics[width=0.5\textwidth]{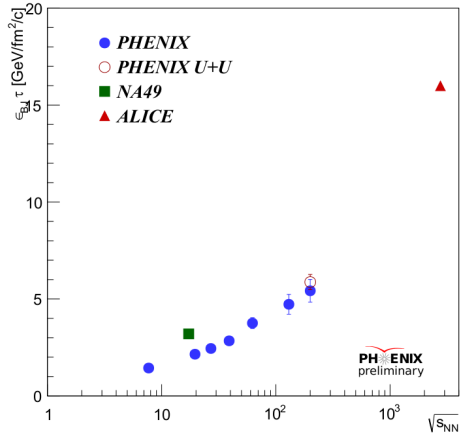}
\end{center}
\caption{The estimated value of the Bjorken energy density, $\epsilon_{BJ}$, multiplied by the formation time in central Au+Au collisions at mid-rapidity as a function of $\sqrt{s_{NN}}$. The open circle represents the estimate for 200 GeV U+U collisions.}
\label{fig:ebjExcite}
\end{figure}

\begin{figure}[htbp]
\begin{center}
\includegraphics[width=0.5\textwidth]{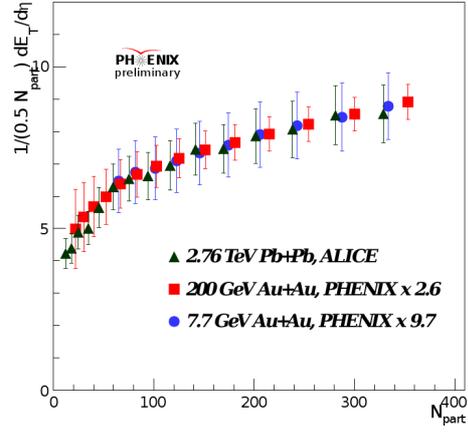}
\end{center}
\caption{$dE_{T}/d\eta$ normalized by the number of participant pairs as a function $N_{part}$. Overlayed are the distributions from 7.7 GeV, 200 GeV, and 2.76 TeV Au+Au collisions. The PHENIX data has been scaled up to overlay the ALICE data \cite{aliceEt}.}
\label{fig:detdetaRHICLHC}
\end{figure}

PHENIX has measured transverse energy ($E_{T}$) production in Au+Au collisions at the following collision energies: 200, 62.4, 39, 27, 19.6, and 7.7 GeV.  These observables are closely related to the geometry of the system and are fundamental measurements necessary to understand the global properties of the collision. This work extends the previous PHENIX measurements in 200, 130, and 19.6 GeV Au+Au collisions \cite{ppg019}. Total $E_{T}$ production results are summarized in Figure \ref{fig:ebjExcite}, which shows the excitation function of the estimated value of the Bjorken energy density \cite{bjorken} expressed as
\begin{equation}
  \epsilon_{BJ} = \frac{1}{A_{\perp} \tau} \frac{dE_{T}}{dy},
\end{equation}
where $\tau$ is the formation time and $A_{\perp}$ is the transverse overlap area of the nuclei. The Bjorken energy density increases monotonically over the range of the RHIC beam energy scan. Also shown is the estimate for 200 GeV U+U collisions taken during the 2012 running period. Although $E_{T}$ production dramatically increases at LHC energies compared to RHIC energies, the shape of the distributions as a function of the number of participants, $N_{part}$, is independent of the collision energy. This is illustrated in Figure \ref{fig:detdetaRHICLHC}, which shows an overlay of the distributions of $dE_{T}/d\eta$ normalized by the number of participant pairs for 7.7 GeV, 200 GeV, and 2.76 TeV Au+Au collisions. The 200 GeV and 7.7 GeV distributions have been scaled up by a factors of 2.6 and 9.7, respectively. The shape of the distribution as a function of $N_{part}$ appears to be driven by the collision geometry.

\section{Charged Particle Multiplicity Fluctuations}

\begin{figure}[htbp]
\begin{center}
\includegraphics[width=0.5\textwidth]{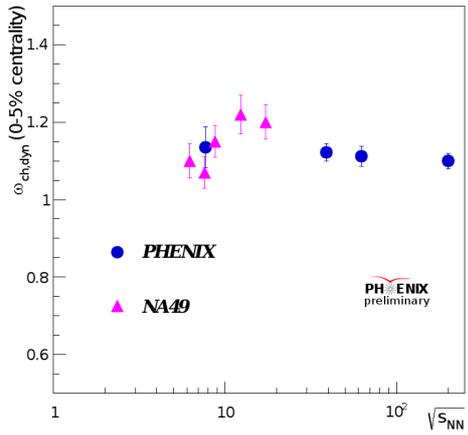}
\end{center}
\caption{Charged particle multiplicity fluctuations in central Au+Au (PHENIX) and Pb+Pb (NA49) collisions expressed in terms of the scaled variance as a function of $\sqrt{s_{NN}}$. Circles represent the results from PHENIX.}
\label{fig:svarExcite}
\end{figure}

Near the QCD critical point, it is expected that fluctuations in the charged particle multiplicity will increase \cite{Stephanov}. PHENIX has extended the previous analysis of multiplicity fluctuations in 200 and 62.4 GeV Au+Au collisions \cite{ppg070} to 39 and 7.7 GeV Au+Au collisions.  Charged particle multiplicity fluctuations are measured using the scaled variance, $\omega_{ch} = \sigma_{ch} / \mu_{ch}$, which is the standard deviation scaled by the mean of the distribution. The scaled variance is corrected for contributions due to non-dynamic impact parameter fluctuations using the method described in \cite{ppg070}. Figure \ref{fig:svarExcite} shows the PHENIX results for central collisions compared to results from the NA49 Collaboration \cite{na49MF} as a function of $\sqrt{s_{NN}}$. There is no indication of the presence of a critical point from the PHENIX results alone.

\section{Higher Moments of Net Charge Distributions}

\begin{figure}[htbp]
\begin{center}
\includegraphics[width=0.5\textwidth]{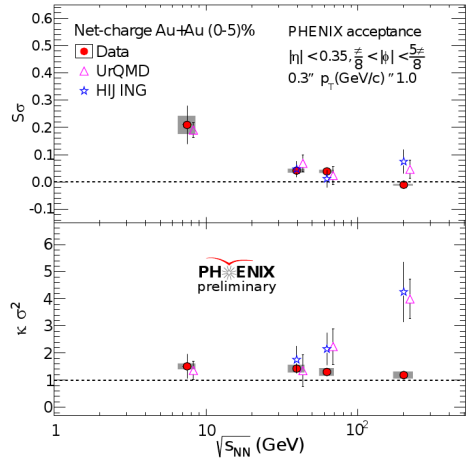}
\end{center}
\caption{The skewness multiplied by the standard deviation and the kurtosis multiplied by the variance from net charge distributions from central Au+Au collisions. The circles represent the data. The grey error bars represent the systematic errors. Also shown are URQMD and HIJING simulation results processed through the PHENIX acceptance. The increase in the kurtosis from URQMD and HIJING may be due to an increase in resonance production at 200 GeV. The blue curve is the result from the Hadron Resonance Gas model \cite{Karsch}.}
\label{fig:skewkurt}
\end{figure}

The shapes of the distributions of the event-by-event net charge are expected to be sensitive to the presence of the critical point \cite{Gavai}. PHENIX has measured the skewness ($S=\langle(N-\langle N \rangle)^{3}\rangle/\sigma^{3}$) and the kurtosis ($\kappa=\langle(N-\langle N \rangle)^{4}\rangle/\sigma^{4} - 3$) of net charge distributions in Au+Au collisions at 200, 62.4, 39, and 7.7 GeV.  These values are expressed in terms that can be associated with the quark number susceptibilities, $\chi$: $S\sigma \approx \chi^{(3)}/\chi^{(2)}$ and $\kappa\sigma^{2} \approx \chi^{(4)}/\chi^{(2)}$ \cite{Karsch}.  The skewness and kurtosis for central collisions are shown in Figure \ref{fig:skewkurt} as a function of $\sqrt{s_{NN}}$. The data are compared to URQMD and HIJING simulation results processed through the PHENIX acceptance and detector response. There is no excess above the simulation results observed in the data at these four collision energies.

\section{Summary}

Presented here are some of the PHENIX results from the RHIC beam energy scan program. From the analyses completed to date, there is no significant indication of the presence of the QCD critical point.  However, many analyses from PHENIX, particularly at $\sqrt{s_{NN}} = $ 27 GeV and 19.6 GeV, will be available soon.



\begin{thebibliography}{00} 
\bibitem{ppg019} S.S.~Adler et al., Phys. Rev. C 71, 034908 (2005).
\bibitem{bjorken} J.~D.~Bjorken, Phys. Rev. D 27, 140 (1983).
\bibitem{aliceEt} C.~Loizides et al., arXiv:1106.6324v1 (2011).
\bibitem{Stephanov} M.~Stephanov et al, Phys. Rev. D 60, 114028 (1999).
\bibitem{ppg070} A.~Adare et al, Phys. Rev. C 78, 044902 (2008).
\bibitem{na49MF} C.~Alt et al, Phys. Rev. C 78, 034914 (2008).
\bibitem{Gavai} R.~V.~Gavai and S.~Gupta, Phys. Lett. B 696, 459 (2011).
\bibitem{Karsch} F.~Karsch and K.~Redlich, Phys. Lett. B 695, 136 (2011).

\end{thebibliography}
\end{document}